# QCD NUCLEAR FACTOR AND THE MOMENTS OF THE MULTIPLICITY DISTRIBUTIONS IN HIGH-ORDER PERTURBATIVE GLUODYNAMICS

A. V. Vinogradov

*Lebedev Physical Institute, RAS, 119991, Leninsky Prospect 53, Moscow, Russia*

**Abstract**

The equation of gluodynamics with modified integral kernel is solved up to 2NLO (next-to-next-to-leading order) and 3NLO (next-to-next-to-next-to-leading order) in perturbative QCD. A relationship between the nuclear factor $N_s$ and the ratio of the cumulant and factorial moments of the multiplicity distribution is examined. A comparison of theoretical results and experimental data in p-Pb and Pb-Pb collisions at 200A GeV and 546A GeV is carried out.

**Introduction**

A process of the multiple particle production is of the greatest importance in high-energy particle interactions. To describe it, a system of coupled QCD equations has been elaborated [1-3]. In case of the running coupling constant, these equations can be solved only within the perturbative QCD [4, 5]. Usually one starts with a more simple case of gluodynamics, i. e. setting aside the quark interactions. It gives a chance to draw some conclusions about behavior of the moments of multiplicity distributions in gluon jets. Taking quarks into account does not significantly affect the results and leaves their qualitative features unaltered [6, 7]. Therefore the case of gluodynamics is considered in the paper.

### I. 2NLO and 3NLO perturbative solutions of the gluodynamics equation

The following equation describes gluon interactions without quarks:

$$G'(y) = \int_0^1 dx K(x)\gamma_0^2(y + \ln x + \ln(1-x))[G(y + \ln x)G(y + \ln(1-x)) - G(y)].$$

Here $G(y,z) = \sum_{n=0}^{\infty}(1+z^n)P_n$ – generating function of the multiplicity distribution, $P_n$ is parton multiplicity distribution in gluon jet, $z$ – auxiliary variable; $G'(y) = dG/dy$, $y = \ln(p\theta/Q_0) = \ln(2Q/Q_0)$ – jet evolution variable, which determines the energy scale, $p$ is initial parton momentum, $\theta$ – the angle of divergence of jet (assumed to be small and constant), $Q$ – the jet virtuality squared four-dimensional momentum, $Q_0$ = const; $\gamma_0^2(y) = \dfrac{6\alpha_s(y)}{\pi}$, $\alpha_S(y)$ is running coupling constant, which is written as $\alpha_S(y) = \dfrac{2\pi}{11y}(1 - \dfrac{51}{121}\dfrac{\ln(2y)}{y}) + O(y^{-3})$ in two-loop approximation.

The moments of multiplicity distribution are defined as $F_q = \dfrac{\sum_{n=0}^{\infty} n(n-1)...(n-q+1)P_n}{\langle n(y)\rangle^q} = \dfrac{1}{\langle n(y)\rangle^q}\dfrac{d^q G(z)}{dz^q}\Big|_{z=0}$ – factorial, and $K_q = \dfrac{1}{\langle n(y)\rangle^q}\dfrac{d^q \ln G(z)}{dz^q}\Big|_{z=0}$ – cumulant moment of q rank; $\langle n(y)\rangle$ - average parton multiplicity of jet with given y.

The kernel of the equation is $K(x) = \dfrac{1}{x} - (1-x)[2 - x(1-x)]$. The nuclear QCD factor $N_s$ usually is inserted into the kernel in the following way [8, 9]: $K(x) = \dfrac{N_S}{x} - (1-x)[2 - x(1-x)]$. It accentuates the change of ratio between the soft and hard parts of parton spectrum in the nuclear media. The singular part $1/x$ describes low-energy particles, and when the $N_s > 1$ emerges in numerator, the soft part of spectrum is spreading, i. e. the number of soft emitted gluons increases. It simulates the softening of the spectrum of produced particles due to their rescattering in the nuclear media. In our case it is convenient to choose the kernel as $K(x) = \dfrac{1}{x} - \dfrac{1}{N_S}(1-x)[2 - x(1-x)]$. It doesn't change the physical interpretation of the phenomenon, but allows us to avoid rising of the coefficient $\sqrt{N_S}$ by the leading term of perturbative series.

The dependence of mean multiplicity on energy is defined by QCD anomalous dimension as $\langle n(y)\rangle = e^{\int^y \gamma(y')dy'}$. The lower limit of integration is constant and influences just the normalization factor, which is free adjustable parameter. That is why the renormalization of $K(x)$ doesn't affect the result. For the QCD anomalous dimension $\gamma$ we use approximate expression $\gamma = \gamma_0(1 - a_1\gamma_0 - a_2\gamma_0^2 - a_3\gamma_0^3 - ...)$ – high-energy (i. e. for large y) perturbative series.

To solve the equation in 2NLO approximation, one should expand each function with shifted argument in Taylor series and keep the terms up to $\gamma_0^2$-order in the right side of equation:

$$\dfrac{1}{\gamma_0^2}[\ln G(y)]'' = G(y) - 1 - 2h_1 G'(y) + h_2 G''(y) + 4Bh_1\gamma_0^2[G(y) - 1].$$

Here $h_1 = \dfrac{11}{24N_S}$, $h_2 = \dfrac{67}{36N_S} - \dfrac{\pi^2}{6}$, $B = \dfrac{11}{24}$ [10].

From the last equation we get coefficients $a_1$ and $a_2$, comparing the terms by the $z$ by first power: $a_1 = h_1 - \dfrac{B}{2}$, $a_2 = (\dfrac{a_1}{2} - h_1)(a_1 + 2B) - \dfrac{h_2}{2}$. Further, examining the $z^q$-terms with arbitrary $q$, the explicit expression for $H_q$ can be obtained:

$$H_q = \dfrac{1 - 2h_1 q\gamma_0 + (h_2 q^2 + 2h_1 a_1 q + 4Bh_1)\gamma_0^2}{q^2 - (2a_1 q^2 + Bq)\gamma_0 + [(a_1^2 - 2a_2)q^2 + 2Ba_1 q]\gamma_0^2}.$$

This function has one minimum, located at $q_{min} \approx \dfrac{1}{h_1\gamma_0} + \dfrac{a_1}{h_1}$, and asymptotically tends to constant $h_2\gamma_0^2$ for large q. When $N_s$ increases, the first root of $H_q$ moves to the left (in the direction of decreasing $q$), while the minimum position moves to the right. Fig. 1 shows the shift of the first root of $H_q$.

In the 3NLO approximation (keeping the $\gamma_0^3$-order terms) we come to the following equation:

$$\dfrac{1}{\gamma_0^2}[\ln G(y)]'' = G(y) - 1 - 2h_1 G'(y) + h_2 G''(y) + 4Bh_1\gamma_0^2[G(y) - 1] + \dfrac{1}{2}h_3 G'''(y) +$$
$$+ h_{12}[G''(y)(\ln G(y))' + G'(y)(\ln G(y))''] - 4Bh_2\gamma_0^2 G'(y),$$

where $h_3 = 2\zeta(3) - \dfrac{413}{108N_S}$, $h_{12} = \zeta(3) - \dfrac{395}{216N_S} + \dfrac{11\pi^2}{72N_S}$ ($\zeta$ – Riemann zeta function) [11].

This equation allows us to find the $a_3$ coefficient ($a_3 = a_2(a_1 - h_1 + \dfrac{3}{2}B) - \dfrac{1}{2}BB_1 + (a_1 + \dfrac{5}{2}B)h_2 - \dfrac{1}{4}h_3$, $B_1 = \dfrac{17}{44}$) and gives a recurrence relation for calculation high-order $H_q$ values via lower $H_q$'s using the well-known formula for cumulant and factorial moments $F_q = \sum_{k=0}^{q-1} C_{q-1}^k K_{q-k} F_k$. The graphs of $H_q$, plotted for $N_s = 1,0..1,08$ (fig. 2) and $N_s = 1,0..1,04$ (fig. 3), demonstrate that when $N_s$ is growing, the first root of $H_q$ now moves to the right and amplitude of $H_q$ oscillations decreases. For $N_s > 1,2$ disappearance of the oscillations may be observed, and behavior of $H_q$ becomes the same as for 2NLO: one minimum and presence of asymptote.

## Comparison with simulated data

In order to ascertain which approximation better corresponds with experimental data, the simulated multiplicity distributions for nuclear collisions of p-Pb and Pb-Pb at the 200A GeV and 546A GeV energy were examined. Using them, the $H_q$ values were calculated and their graphs were plotted (fig. 4, fig. 5). One can see that the behavior of $H_q$ is more precisely fitted by

3NLO approximation, since the first root of $H_q$ moves to the left when $N_s$ is growing, and amplitude of oscillations decreases. Comparison of fig. 3 and fig. 4 permits to estimate the value of $N_s$ for Pb-Pb collision: $N_s \approx 1,0$ for 200A GeV and $N_s \approx 1,04$ for 546A GeV.

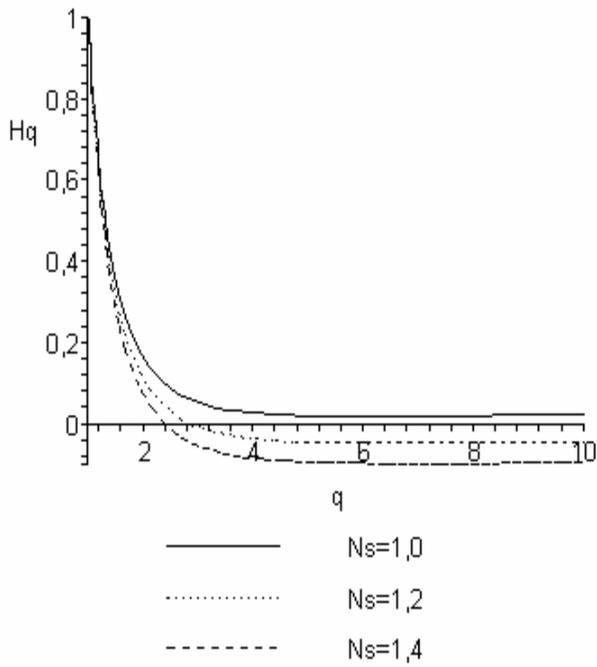

Fig. 1. Shift of the roots of $H_q$ according to increasing $N_s$ in the 2NLO approximation.

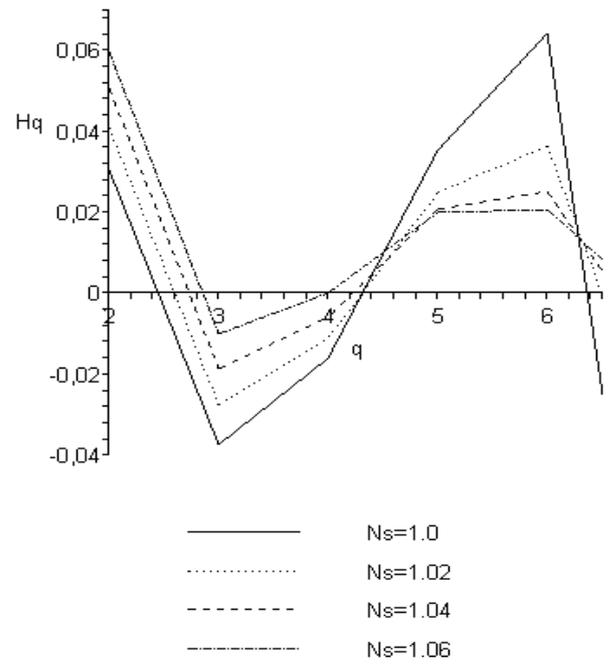

Fig. 2. Shift of the roots of $H_q$ according to increasing $N_s$ in the 3NLO approximation.

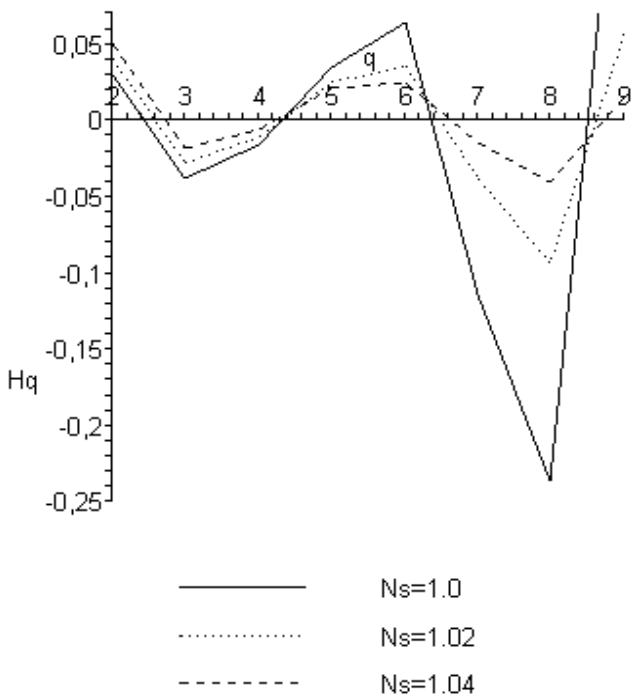

Fig. 3. Behavior of $H_q$ when $N_s$ is growing in the 3NLO approximation.

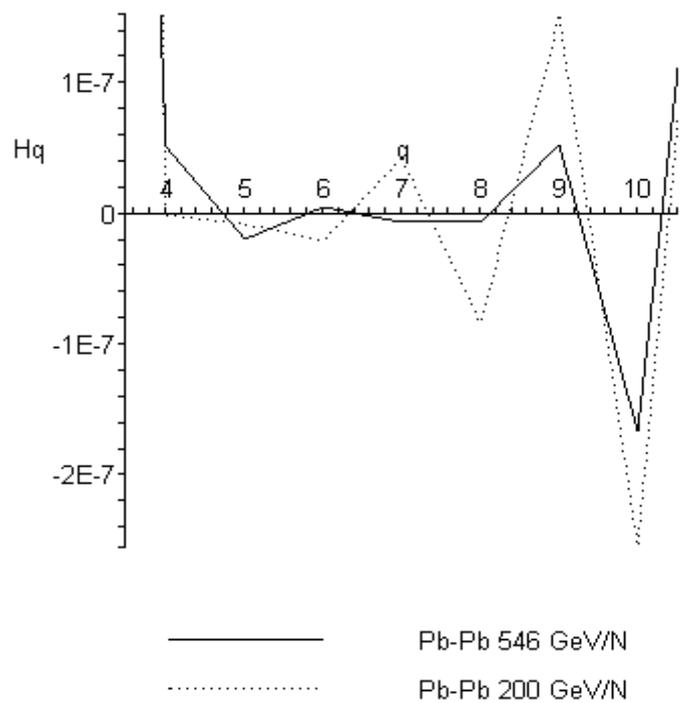

Fig. 4. $H_q$ for the collisions of Pb-Pb nuclei with energies 200A GeV and 546A GeV.

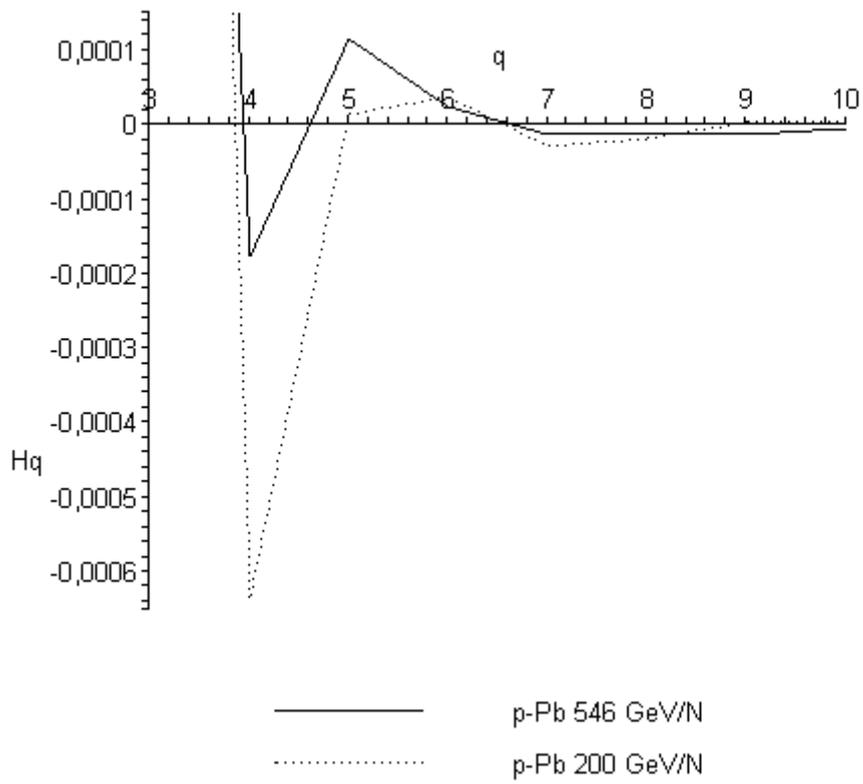

Fig. 5. $H_q$ for the collisions of p-Pb nuclei with energies 200A GeV and 546A GeV.